# Strain-Based Room-Temperature Non-Volatile MoTe$_2$ Ferroelectric Phase Change Transistor

Wenhui Hou[1], Ahmad Azizimanesh[1], Arfan Sewaket[1], Tara Peña[1], Carla Watson[2], Ming Liu[3], Hesam Askari[4] & Stephen M. Wu[1,2]

[1]*Department of Electrical and Computer Engineering, University of Rochester, Rochester, NY 14627, USA*

[2]*Department of Physics and Astronomy, University of Rochester, Rochester, NY 14627, USA*

[3]*Electronic Materials Research Laboratory, Key Laboratory of the Ministry of Education and International Center for Dielectric Research, Xi'an Jiaotong University, Xi'an 710049, Shaanxi, China*

[4]*Department of Mechanical Engineering, University of Rochester, Rochester, NY 14627, USA*

*email: stephen.wu@rochester.edu

**Abstract summary**

**The primary mechanism of operation of almost all transistors today relies on electric-field effect in a semiconducting channel to tune its conductivity from the conducting 'on'-state to a non-conducting 'off'-state. As transistors continue to scale down to increase computational performance, physical limitations from nanoscale field-effect operation begin to cause undesirable current leakage that is detrimental to the continued advancement of computing[1,2]. Using a fundamentally different mechanism of operation, we show that through nanoscale strain engineering with thin films and ferroelectrics (FEs) the transition metal dichalcogenide (TMDC) MoTe$_2$ can be reversibly switched with electric-field induced strain between the 1T'-MoTe$_2$ (semimetallic) phase to a semiconducting MoTe$_2$ phase in a field effect transistor geometry. This alternative mechanism for transistor switching sidesteps all the static and dynamic power consumption problems in conventional field-effect transistors (FETs)[3,4]. Using strain, we achieve large non-volatile changes in channel conductivity ($G_{on}/G_{off}\sim10^7$ vs. $G_{on}/G_{off}\sim0.04$ in the control device) at room temperature. Ferroelectric devices offer the potential to reach sub-ns nonvolatile strain switching at the attojoule/bit level [5–7], having immediate applications in ultra-fast low-power non-volatile logic and memory[8] while also transforming the landscape of computational architectures since conventional power, speed, and volatility considerations for microelectronics may no longer exist.**

We design our device using single crystal oxide substrates of relaxor ferroelectric Pb(Mg$_{1/3}$Nb$_{2/3}$)$_{0.71}$Ti$_{0.29}$O$_3$ (PMN-PT) as the gate dielectric (0.25-0.3 mm thickness). On top of this ferroelectric substrate we exfoliate 1T'-MoTe$_2$ (13-70 nm) from a single crystal source, and pattern devices using Ni contact pads (Figure 1a,b). Exfoliation was performed in a humidity-controlled environment for increased adhesion. Depending on the contact material, stress from the deposited thin film strains the MoTe$_2$ channel at the contact pads, analogous to the uniaxial strain techniques from strained silicon technology, widely adopted in industrial CMOS processes[9]. We find that contact metal stress and TMDC to substrate adhesion are both **critically** important for obtaining a functional device. During fabrication we are careful not to increase the temperature of the ferroelectric above 80° C, well below the Curie temperature of the ferroelectric at 135° C. Upon reaching the Curie temperature, the sudden quenching through the transition will cause the size of the domains to shrink from the few micron scale to the nanometer scale[10], setting a complicated strain state within the MoTe$_2$ (see Figure S1-S2).

We choose to characterize the strain in the system by micropatterning directional strain gauges on the ferroelectric surface and to characterize the electrical properties of the MoTe$_2$ device using standard transfer curve measurements (Figure 1a). Dynamic strain from the ferroelectric is measured through strain gauges patterned using the same thin film Ni as the contact pads to the MoTe$_2$ channel. Therefore, changes in strain in the gauge, as the ferroelectric is poled, represents an experimental measure of changes in thin film stress applied to MoTe$_2$ at the contact pads by the contact metal. Strain starts off bipolar with electric field, typical for ferroelectrics, then upon further cycling develops an asymmetry that causes a transition to unipolar (non-volatile) operation. This asymmetry is independent of crystal orientation of the ferroelectric and the direction of the gauge. We attribute the asymmetry in the strain curves to the well-known effect of elastic dipoles in ferroelectric systems causing an internal electric field bias effect[11,12], which are aligned due to thin film stress from Ni as the thin film is worked[13] (see supplemental discussion and Figure S15). Since the size of the strain gauges are large relative to the individual domain structure of the ferroelectric (100 μm vs. 1 μm) we see an aggregate strain effect, where individual MoTe$_2$ flakes may feel domain dependent strain[14].

To test the effect of strain on our devices, we pattern an exfoliated flake of 13 nm 1T'-MoTe$_2$ with 35 nm Ni contact pads, which applies a measured in-plane tensile stress of 0.58 GPa to MoTe$_2$ at the contacts (see Table S1). The measured transfer characteristics on a linear scale are presented in figure 1d. We see a reversible **room-temperature** on-off switching behavior that matches the standard strain butterfly curve in ferroelectric materials (Figure 1c). We observed substantial changes (> 1 order of magnitude) in channel conductivity in over 10 other MoTe$_2$ devices of various thicknesses on various PMN-PT (011) substrates and observed similar bipolar effects. Several devices required multiple sweeps of gate electric field before the strain driven conductivity changes occurred but repeated for several cycles afterwards in a stable state after training. This training behavior is likely due to enhanced adhesion between the MoTe$_2$ and the ferroelectric upon cycling[15]. Variation in adhesion to rough substrates may contribute to some of the observed differences in training (see Figure S12). The bipolar nature of channel current with respect to electric field strongly suggests a strain driven phase transition between semimetallic and semiconducting phases of MoTe$_2$, where the strain in the MoTe$_2$ flakes evolves with applied gate voltage across the ferroelectric. We note that in our subsequent testing, MoTe$_2$ flake thickness was not significantly correlated with differences in device behavior once a phase change has occurred but may still influence whether phase changes can occur at all due to insufficient strain transfer. We choose to focus on devices that have shown clear changes in phase as measured through electrical characterization.

To prove that this behavior results in a transition to a semiconducting phase, we measure current-voltage characteristics in the low-conductivity state of a representative device of the same geometry and observe a double Schottky junction between the Ni contacts and the phase-transformed MoTe$_2$ channel (Figure 1e). We note that in our process only 1T'-MoTe$_2$ is used and thus no material with a gap exists within our transistor structure as fabricated; the Schottky-like behavior only appears after a transition to the low-conductivity state as a result of strain-driven switching.

To further examine the phase transition in these devices, we performed temperature dependent measurements of channel conductivity in another MoTe$_2$ device with a nominal thickness of 70 nm, using a PMN-PT (011) substrate with Ni contact electrodes. An optical micrograph of the actual measured device is presented in figure 1b. Log-scale conductivity is shown in figure 2, showing a unique temperature evolution as we sweep from 300 K to 270 K to 330 K and back to 300 K. Both bipolar and unipolar (non-volatile) channel conductivity modulation was observed, with a maximum $G_{on}/G_{off}$ ~

$6.2 \times 10^6$ in the final 300 K state, which is a larger value than any 2H-MoTe$_2$ field-effect transistor using any contact scheme for any thickness[16–18]. For a channel thickness of 70 nm, representing ~100 layers of MoTe$_2$, conventional field effect conductivity modulation is limited to less than 1 order of magnitude due to electric field screening in the semiconductor[19]. By purposely choosing a contact metal that exhibits Schottky contact behavior with semiconducting MoTe$_2$, our 'off-state' becomes two back-to-back Schottky diodes exhibiting low current when bias voltages are kept low.

Since MoTe$_2$ flakes may land on any single ferroelectric domain within the PMN-PT domain structure, it is understandable why device to device variation may occur since strain directionality may play a large role in seeding the phase transition[14,20]. These uncontrolled factors lead to variation in bipolar modulation behavior, with figure 2 showing the opposite result of figure 1. The unique temperature hysteresis within this device can be understood from the perspective that the phase boundary with respect to strain is highly temperature dependent (see Figure S15), which has been measured in the past in 2H-MoTe$_2$ suggesting the 2H phase is more favorable at low temperatures[21]. As measured through our strain gauges, continued electric field sweeping of the ferroelectric causes the electrical characteristics of our device to transition from bipolar to unipolar operation, as the strain from the Ni contacts change and eventually reach a stable unipolar strain after repeated cycling. The majority of device behavior from figure 2, can therefore be explained by noting that the semiconducting phase is more stable at low temperature, while the semimetallic phase is more stable at high temperature, and that the variation in transfer characteristics with electric-field sweeping is due to variation in strain applied to the MoTe$_2$ from the Ni contacts. There are other uncontrolled factors that may also contribute to a smaller degree, including differential thermal contraction, thin film stress relaxation with respect to temperature, and temperature dependence of the piezoelectric coefficient in PMN-PT. We note that with further engineering beyond our initial presentation here, it is possible to obtain **both** unipolar or bipolar behavior through engineering ferroelectric strain from: selective electric field sweeping[22], ferroelastic domain switching[23], or internal electric-field bias engineering[24].

To directly view a real space image of the channel under ferroelectric strain, we use conductive AFM (CAFM) to directly probe channel conductivity. Figure 3 shows the results of CAFM scans of the device presented in figure 2 directly after measurement (left in the low conductivity state) with both contact pads grounded with respect to the voltage biased CAFM tip. A large non-conductive area is found near the contact edge of the device representing the semiconducting phase of MoTe$_2$ and hinting at effects arising from contact metal induced strain. We further examine the effect of strain on this channel by applying a gate voltage pulse to set the conductivity state of the channel as schematically represented by the hysteresis loops in figure 3c-e. We first pulse the gate in the same direction that the device is already set in, as a control measurement, and find in figure 3d that the large non-conducting region at the contact edge is retained, but is now also mirrored at the other contact edge as well. The initial asymmetry of the CAFM image (Figure 3c) is due to gate voltage being applied with respect to the source contact from the measurement in figure 2, whereas the first and second pulse (Figure 3d,e) are applied with respect to both source and drain. Next a pulse is applied in the positive direction to switch the channel to the conducting state and we find in figure 3e that the non-conductive regions near the contacts now close. This suggests that the strain driven phase transition is seeded by strain from the thin film stress induced by the contacts.

We explore the effect of substrate crystal orientation and further examine the effect of contact metals on the behavior of our devices by exploring more MoTe$_2$ devices on PMN-PT (111) substrates. Figure 4a shows the same current map measured through CAFM on a MoTe$_2$/PMN-PT (111) device with the same Ni contacts, showing the same characteristic phase change behavior at the contact edges but on a device with a shorter channel. Our contact strain hypothesis is supported by a finite element analysis simulation of the strain state within a MoTe$_2$ channel given tensile strain from the edges, showing the same characteristic shape as the current map. Further modeling done on a longer channel device, similar to the device presented in figure 3, shows the same correspondence (see Figure S6). With the calculated strain induced into the MoTe$_2$ channel from contact induced stress (using literature values for MoTe$_2$ Young's modulus and Poisson's ratio[21]), and the characteristic size of the semiconducting region as measured by CAFM at the contact edge, we can extract the approximate threshold for the phase transition. We find that the contacts apply tensile strain to the MoTe$_2$ at 0.4%, and the strain threshold occurs at approximately 0.33% based off the length scale that the semiconducting region bleeds out at the Ni contacts on the device from Figure 3. The magnitude of this strain is comparable with both experimentally observed and theoretically predicted strain transitions in MoTe$_2$[20,21,25], as well as with the amount of electric field controllable strain in PMN-PT as measured in figure 1c. Since these devices were patterned on PMN-PT (111), it suggests that phase transitions are robust against ferroelectric orientation. We expect these microscale devices do not depend heavily on the overall aggregate strain behavior of the ferroelectric single crystal, but on the individual ferroelectric domains that the MoTe$_2$ channel and contacts land on[14].

Using different contact metals on PMN-PT (111) we also show that phase transitions are only robust when contact metals apply a finite tensile stress (Ni). When compared to low stress contacts (-0.2 GPa) made of 50 nm Ag, conductivity changes are limited to few percent range at all temperatures compared to a similar Ni device which has conductivity changes ~$10^9$ % (Figure 4b). Out of 13 measured devices with Ag contacts on both PMN-PT (011) and (111) oriented substrates, no device showed any meaningful conductivity modulation other than a marginal few percent change. To further prove the necessity of a tensile stressor layer, an alternative MoTe$_2$ device was also constructed with low-stress Ag contacts encapsulated in a high-stress tensile insulating MgF$_2$ layer as the static stressor. With MgF$_2$ applying a similar magnitude tensile strain as Ni (see Figure S10), the devices again returned to large conductivity switching with PMN-PT (see Figure S11).

The overall predicted mechanism of operation based on our experimental devices is outlined in figure 4a. Thin film stress from contact metal deposition sets a higher strain state than a single ferroelectric can apply by itself, while a small amount of electric-field controllable strain from PMN-PT can bring the MoTe$_2$ across the phase boundary. This suggests that the majority of the channel conductivity changes happen underneath the contacts and our CAFM measurements are only able to incidentally observe phase transitions at the edges in special cases. This is supported by the fact that no CAFM measurement on any device showed channel conductivity changes further than 250-500 nm away from the contact edges. The use of metastable 1T'-MoTe$_2$ as the starting material allows the two contact sides to always be connected through a metallic link, such that large changes that occur near the contacts are reflected in the final electrical measurement. Phase transitions were not observed when using 2H-MoTe$_2$ as the starting material with Ni contacts, although presumably through thin film strain engineering it can become possible in a different geometry. We also note that while the majority of the devices had large changes in conductivity, approximately 8 out of 28 measured devices on both (011) and (111) PMN-PT using Ni contacts also had low modulation in the few percent range [1/3 on (111), 1/4 on (011)] (see Figure S7). We attribute this variation to the various uncontrolled aspects of our devices: whether each MoTe$_2$ flake would land on single or multiple ferroelectric domains, what the polarization of the domain was (setting the zero strain starting state)[14], which direction the ferroelectric strain exists in with respect to the contact metal, and what the crystal orientation of the MoTe$_2$ was when exfoliated.

To further examine the properties of the semiconducting region in our devices after phase changes have occurred, we perform backside Raman spectroscopy on MoTe$_2$ devices with Ni contacts fabricated on transparent double-side polished MgO substrates (0.25 mm thickness) in the same geometry as on PMN-

PT substrates (see Figure S13-S14). By performing Raman measurements **through** the substrate, we are able to examine the phase-changed material underneath the Ni contact. We find that the semiconducting phase in our MoTe$_2$ devices are not the 2H semiconducting phase, and the additional Raman peaks that appear are similar to Raman signatures from other phase-transformed MoTe$_2$ devices from literature (driven by strain[21] or laser heating[16]). Detailed discussion on the exact nature of the structural-phase change in our devices, as well a thorough elimination of potential trivial non-strain-induced phase-change mechanisms are included in the Supplementary Information.

We have endurance tested Ni contact devices and have seen that reliable stable switching behavior only occurs for between 40-70 switches (see Figure S4). Beyond this, the device becomes unstable as a result of the transfer characteristics changing upon further device cycling. A potential source of instability is the thickness of our samples, as strain cannot be fully transferred without eventual slippage between Van der Waals bonded layers. Recent studies have shown that strain can be applied in 2D bonded systems up to 40 nm[26], although the exact nature of strain transfer from the 3D bonded limit to the 2D bonded limit needs to be further explored before the exact nature of interlayer slippage can be quantified[27,28]. For the eventual application of a strain transistor, reducing device thickness down to the monolayer limit will become important to eliminate the variable of layer sliding to further explore other device failure mechanisms and to find the ultimate limit of endurance in our strain-driven devices. Other factors in endurance may include effects such as, improper flake adhesion or plastic deformation of static stressor layers. Devices have also been tested for speed and have been shown to switch with gate voltage pulses of 10 ms. This is the limit of our experimental capability since we switch the entire single crystal ferroelectric over 0.25-0.3 mm, which requires pulses of 150 V. Further testing on thin film ferroelectrics will need to be performed to find the true limit for the speed of strain switching. The ultimate limit is likely similar to existing speeds of MoTe$_2$ phase change memristors that operate down to the ns scale[29], with the potential to go as low as the ps range since ferroelectric switching has been shown to operate in this range[6]. Finally, while PMN-PT and MoTe$_2$ are used in this first demonstration, many other materials have been proposed to be sensitive to strain-based phase change[20,30], and many different ferroelectrics[6] can be used to further engineer strain. Further development of materials for electrical and mechanical properties will become important to engineer for the factors listed above as development becomes more important for this technology.

We have introduced a new type of transistor where electric-field induced strain can reversibly change the device from semimetallic to semiconducting. Strain induced phase changes do not suffer from the same limitations as conventional field effect transistors in terms of obtaining large on-off ratios while retaining fast switching outside of subthreshold slope limitations. The 'on-state' of our device is fully metallic leading to exceptionally high on-currents, while the 'off-state' can be engineered for small current leakage through contact engineering. Since the devices do not heavily depend on the thickness of the MoTe$_2$ channel and retains the three-terminal gate configuration from conventional field effect transistors, the process to scale these devices into realistic commercial integrated circuits becomes significantly less challenging. This type of 'straintronic' device, combines the best properties of 2D materials (large elastic limit, immunity to strain induced breakage, wide variety of phases) with the best properties of ferroelectrics (low-power, non-volatile, fast switching). The performance of the individual constituent elements suggests that there is a much higher ceiling to device performance than presented here in our first demonstration. Looking beyond MoTe$_2$, using strain engineered 2D materials with ferroelectrics represents a fundamentally exciting platform to explore the wide variety of other electric-field induced phase transitions in the 2D materials world (i.e. magnetic[31], topological[32], superconducting[33]...), leading to a new method for low-power control over various quantum and conventional states of matter.

## Acknowledgments

This work made use of the Cornell Center for Materials Research Shared Facilities which are supported through the NSF MRSEC program (DMR-1719875). The authors thank D. H. Kelley for the borrowed usage of his MBraun glovebox, as well as A. Nick Vamivakas and A. Mukherjee for discussions/assistance in micro-Raman spectroscopy.

## Author Contributions

Device fabrication performed by W.H., A.S., T.P., and A.A.; Device characterization performed by W.H., A.A., and S.M.W.; Conductive atomic force microscopy performed by W.H., A.S., and S.M.W.; Strain gauge calibration performed by W.H. and S.M.W.; Topographic atomic force microscopy and optical contrast calibration performed by T.P.; Thin film stress measurements performed by C.W., A.A., W.H., and S.M.W.; Piezoresponse force microscopy performed by C.W.; Raman spectroscopy performed by A.A. and S.M.W.; Finite element analysis simulation performed by H. A.; PMN-PT single crystals provided by M. L.; Original experiment conception and project supervision provided by S.M.W.;

## Competing interests

The authors declare no competing interests.

## Figures

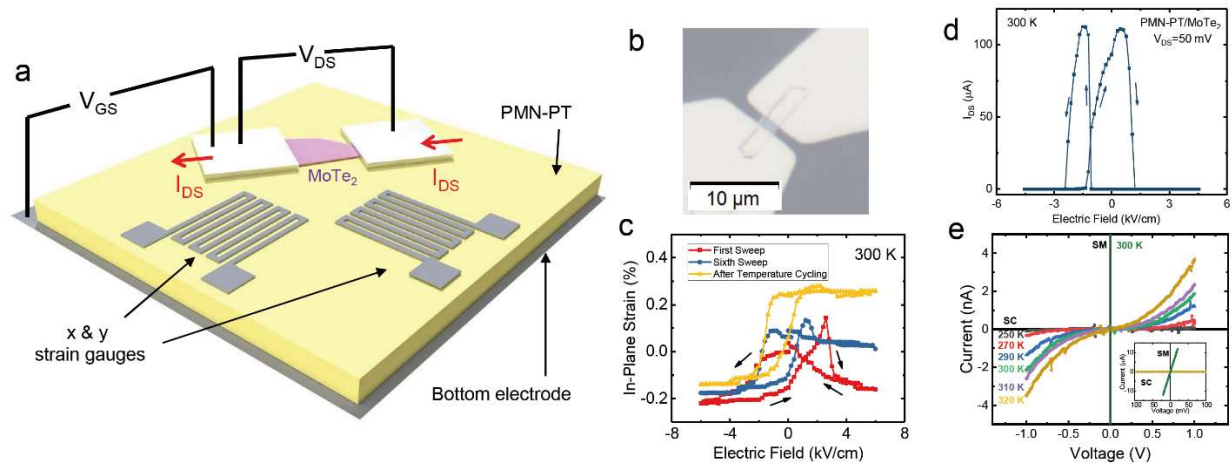

**Fig. 1 | Device Schematic and Operation.** (a) Schematic representation of ferroelectric strain field effect device. (b) Optical micrograph of actual MoTe$_2$ on PMN-PT (011) device used in Figure 2 and 3 is presented. (c) Measured strain curves from as-deposited Ni strain gauges measuring evolution of strain in thin film Ni contacts, after the first sweep, the 6$^{th}$ sweep, and finally after temperature cycling to 250 K to 330 K and back to 300 K. (d) Strain induced transistor operation on a 13 nm MoTe$_2$ channel, with device W/L=2. (e) Current-voltage (IV) characteristics of representative devices in the semimetallic (SM) and semiconducting (SC) states, with the semiconducting state showing Schottky behavior with respect to temperature. Inset shows the same IV curves on an expanded scale to highlight the semimetallic state.

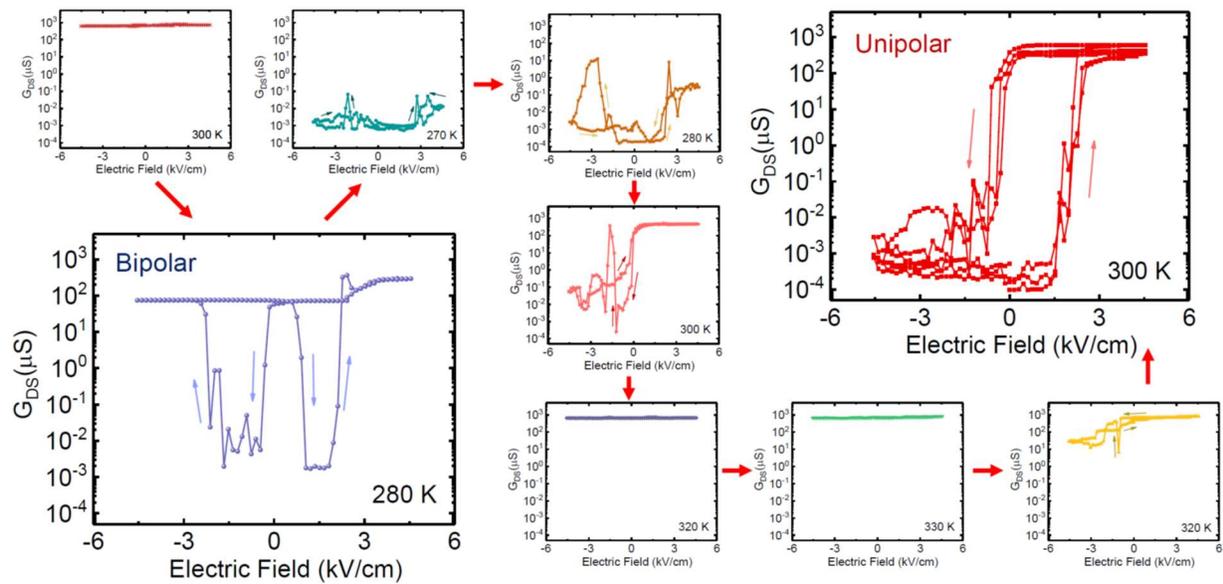

**Fig. 2 | Temperature cycling and non-volatile switching.** Log-scale channel conductance measurements for the device presented in figure 1b (70 nm channel, W/L=2.3) with respect to temperature cycling. Initially starting at 300 K then to 270 K to 330K to 300 K. Measurements are taken with $V_{ds}$= 100 mV and are current limited at 100 μA. Transitions between full semimetallic state, and Schottky barrier limited operation in the semiconducting state are shown. Depending on temperature history, large variations in device behavior exist due to strain evolution and temperature dependence of the phase transition. Both bipolar and unipolar (non-volatile) behavior within the devices are seen. Final unipolar device at a single temperature is robust as shown by three full major loop sweeps. Additional data on temperature dependence, unipolar behavior, and device stability are included in the supplementary information (see Figure S3,S5).

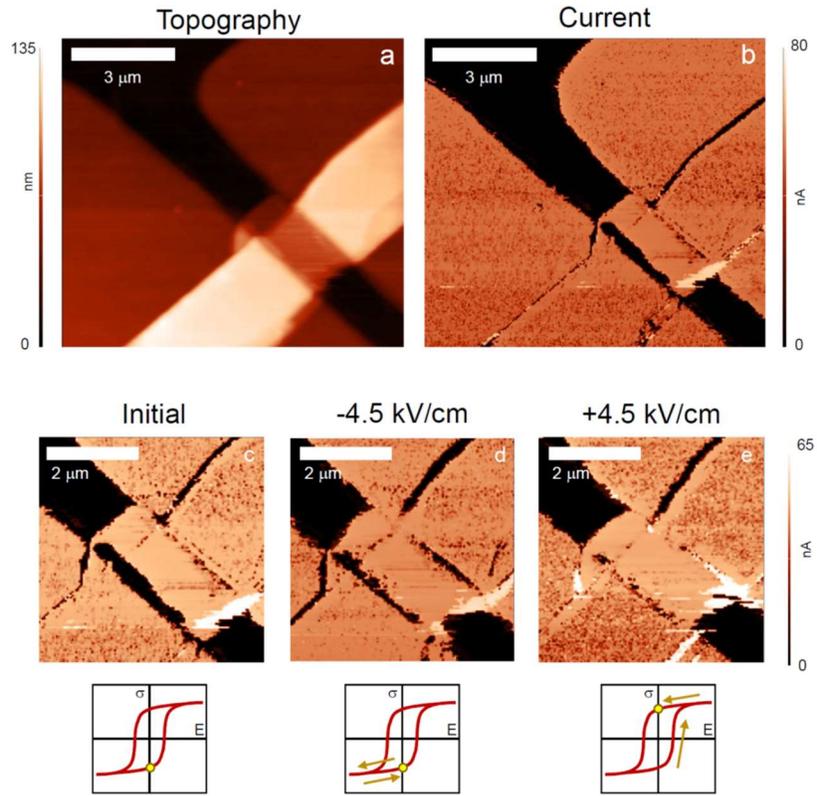

**Fig. 3 | Conductive atomic force microscopy of switching behavior.** (a),(b) Topography and current as measured by conductive atomic force microscopy of the device presented in figure 2 after being left in the low conductivity state (290 K). Large changes in conductivity can be inferred from CAFM data near the contact edges. (c),(d),(e) CAFM images after a pulse sequence representing the initial condition (c), then a pulse in the same direction as polarization (d), and finally a pulse in the opposite direction as polarization (e). Channel near contact edges stay low conductivity for (d), while the edges go higher conductivity after the opposite pulse (e).

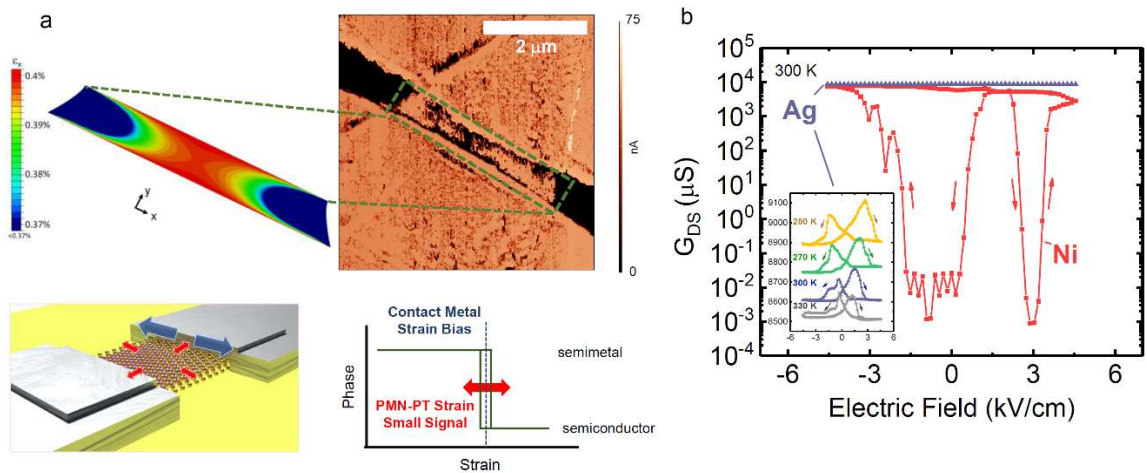

**Fig. 4 | Effect of contact metals and simulations.** (a) CAFM on 50 nm MoTe$_2$ device (W/L=6.3) patterned with Ni contacts on PMN-PT (111) oriented single crystal showing contact edge effects. Data is shown with finite element analysis simulation of strain in the channel assuming clamped tensile strain from the contact metal. Device operation based on contact metal induced strain is presented below, where electric field controllable strain from PMN-PT (red) is used to modulate on top of contact induced strain (blue). Proposed mechanism of operation of strain-biased PMN-PT device is also presented. (b) Channel conductance on MoTe$_2$ devices with Ag (low compressive stress, -0.2 GPa) and Ni (larger tensile stress, 0.58 GPa) contacts. Only small modulation in conductivity is seen in Ag contact devices (~4 %) versus Ni devices (~10$^9$ %). Inset represents a temperature evolution of an Ag contact device.

## Methods

### Device Fabrication

Devices were fabricated on PMN-PT single crystals (0.25-0.3 mm thickness) with sputtered Au (100 nm)/ Ti (5 nm) bottom electrode contacts. Commercially purchased 1T'-MoTe$_2$ (HQ Graphene) was exfoliated onto the polished (R$_a$~0.5 nm) side of PMN-PT using standard Nitto Semiconductor Wafer Tape SWT10. Exfoliation was either performed in a low humidity room (relative humidity < 10%) or in a controlled glove box with < 1 ppm of H$_2$O and O$_2$. Optical contrast thickness identification was used to characterize thickness of flakes (see Figure S8-S9). Direct-write laser photolithography was performed using a Microtech LW405 laserwriter system, with S1805 photoresist that is specifically softbaked at low temperatures (80° C) to prevent heating above the Curie temperature. If standard bake recipes for photoresists and e-beam resists are used, spontaneous quenching through the Curie temperature will occur, and result in devices that do not produce larger than a few percent conductance modulation. Patterns were exposed using standard photolithographic doses of 300 mJ/cm$^2$, and photoresist was soaked in chlorobenzene for 5 minutes before development for undercut control. All contact metals were deposited using e-beam evaporation at $5 \times 10^{-5}$ torr pressure at a rate of 1 Å/s. Strain gauges were constructed from the same thin film deposition (35 nm Ni), and separately calibrated using flexible Kapton substrates with strain applied through bending. Axial and transverse gauge factors are measured to be 3.1 and 0.15 respectively, limiting the contributions of strain perpendicular to the axial direction by over a factor of 20.

### Device Characterization

Devices were characterized using low-frequency AC lock-in techniques (3 Hz) with AC voltage signal provided by a separate phase locked function generator. Measurements of conductivity were performed with applied voltages of 100 mV with a 100 µA current limiting circuit to prevent device blow-out since the high conductivity states are purely metallic and large current densities can form when the phase transition occurs. More detailed measurements of Ag electrode devices were performed with 10 mV applied voltage with the same current limiter, but all devices were first tested under the same conditions as Ni electrode devices to confirm the null result. Gate voltages were applied between the backgate and the source contact in the device using a

DC power supply and typically applied for 5 seconds before each conductivity measurement. All devices were swept in positive electric field and cooled in negative remanence unless otherwise noted.

**Conductive Atomic Force Microscopy**

Devices were measured using conductive tips coated using confocal DC sputtering of 10 nm W, followed by 20 nm Pt. Measurements were performed in contact mode, with force setpoint low to prevent sample damage upon scanning. Pulse measurements were done by removing the device from the AFM, ramping voltage on the gate relative to both grounded contact pads over 30s and then ramping down to 0 V. Devices were then placed back into the AFM for the next CAFM measurement.

**Finite Element Analysis**

Finite element analysis was performed using Abaqus FEA software suite. A membrane with the same size of the thin film was modeled by quadratic plane elements, with average side length of 0.05 μm. Lateral strain of 0.4% was applied to the contact edges to model Ni interface strains. Side edges are free. The material is assumed to be isotropic by averaging anisotropic mechanical properties. Color coding of resulting strain contours are set to show the range where switching is expected based on the strain analysis described on the main text.

**Raman Spectroscopy**

Raman spectroscopy was performed using a Renishaw InVia Confocal Raman Microscope, with a 50x long working distance objective to image through the transparent double side polished ($R_a$~0.5 nm) MgO substrate (532 nm excitation laser). Laser power was limited to 2 mW to limit changes to the device due to heating. Devices were electrically confirmed to be in the semiconducting state through Schottky barrier IV measurements, then located in the Raman microscope through the backside to gain access to the $MoTe_2$ under the contacts. The estimated spot size of the laser was 1.5 μm.

# Supplementary Information

**Supplementary Figures**

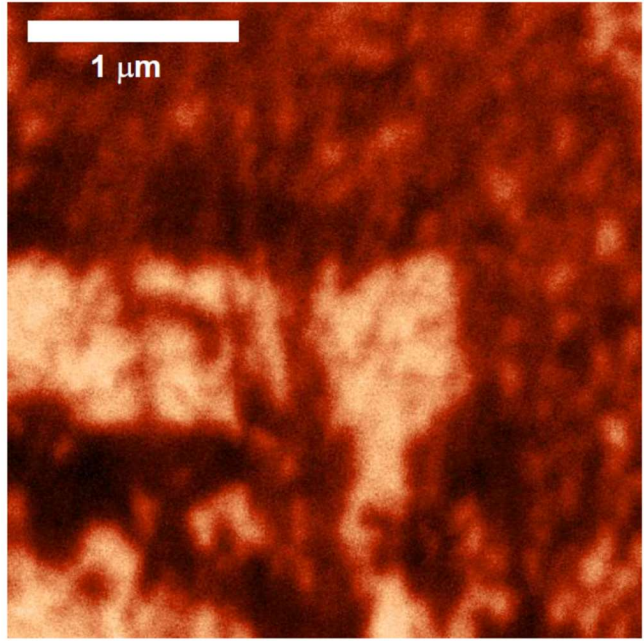

**Supplementary Figure 1.** Piezoresponse force microscopy (PFM) on a section of PMN-PT (011) after quenching through the Curie temperature from 150° C with 10 $V_{pp}$ excitation on-resonance at 280 kHz. Out-of-plane contrast in PFM phase shows formation of nanoscale ferroelectric domains within larger micron sized domains. The formation of these nanodomains is consistent with the domain evolution of PMN-PT single crystals from literature[1,2] and leads to a non-uniform strain state within $MoTe_2$ devices. Common device processing techniques typically require heating near the Curie temperature, which we avoid by using a low temperature soft-bake with standard photolithographic photoresists.

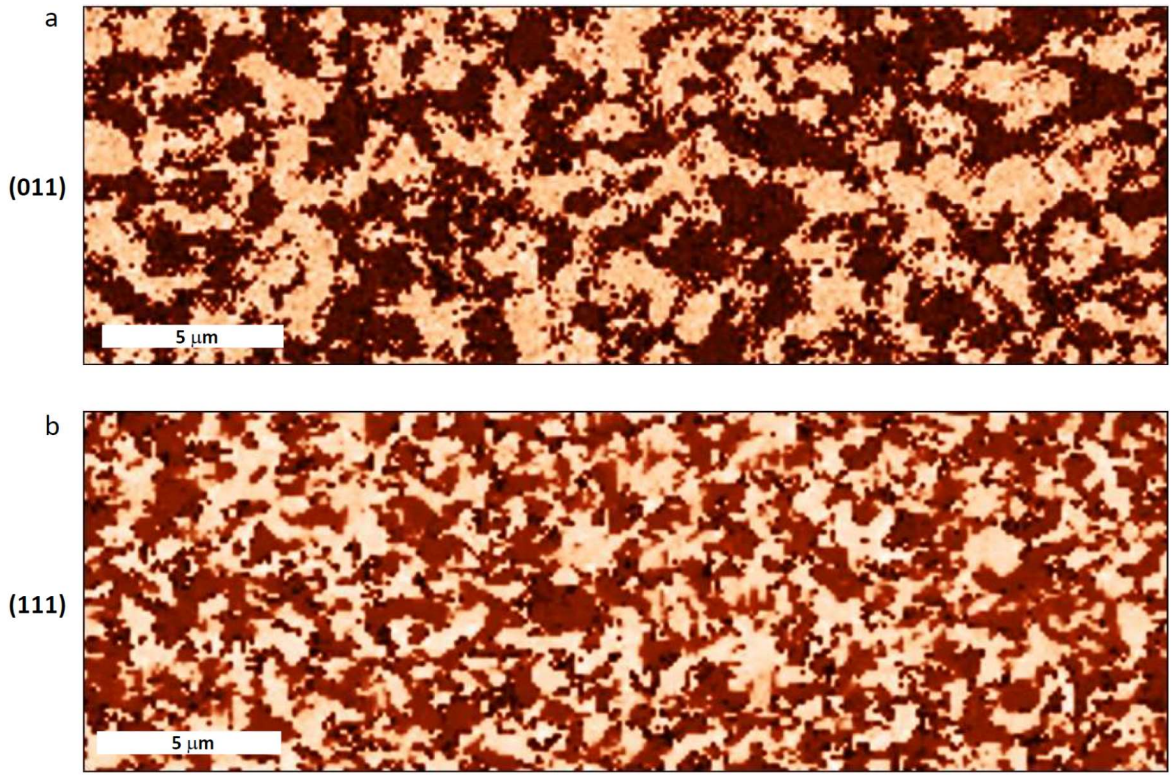

**Supplementary Figure 2.** PFM contrast image of ferroelectric domain structure of two representative PMN-PT samples used in our work, for (a) (011) and (b) (111) oriented single crystal substrates before quenching. Domain size is roughly in the few micron range, on the typical order of our device size.

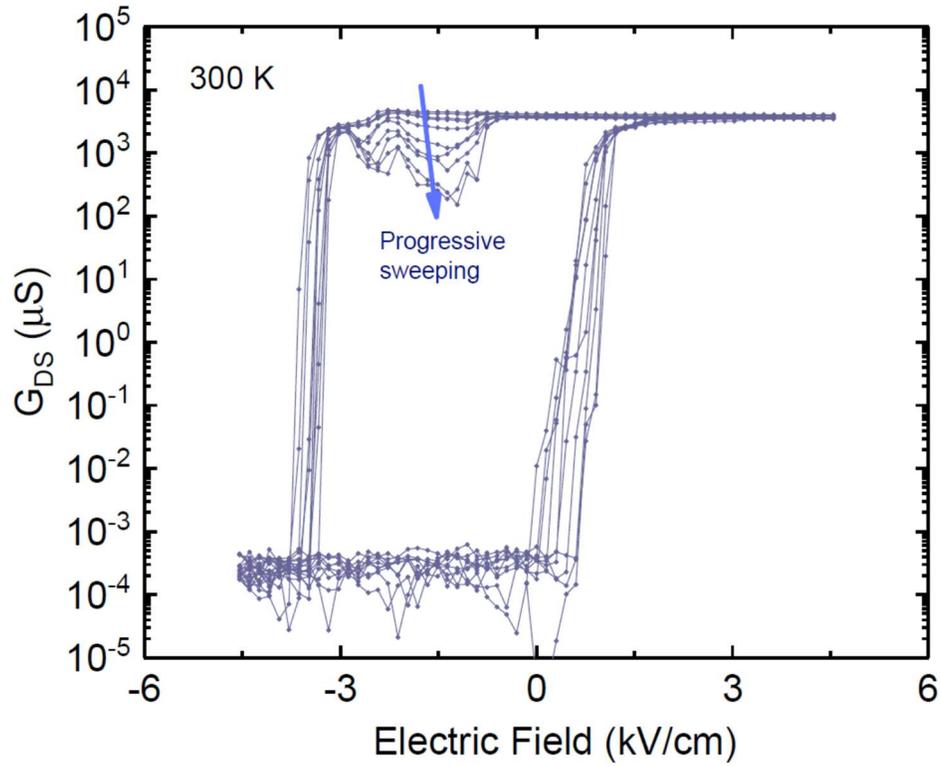

**Supplementary Figure 3.** Unipolar switching behavior on a MoTe$_2$/PMN-PT (011) device (17 nm MoTe$_2$ thickness, W/L=3.25), separate to the data presented in the main paper, showing ten full gate voltage sweeps. Behavior is after temperature sweep from 250 K to 325 K and back to 300 K. With progressive sweeping, a dip within the transfer characteristics begins to grow, likely due to the evolution of strain within this device.

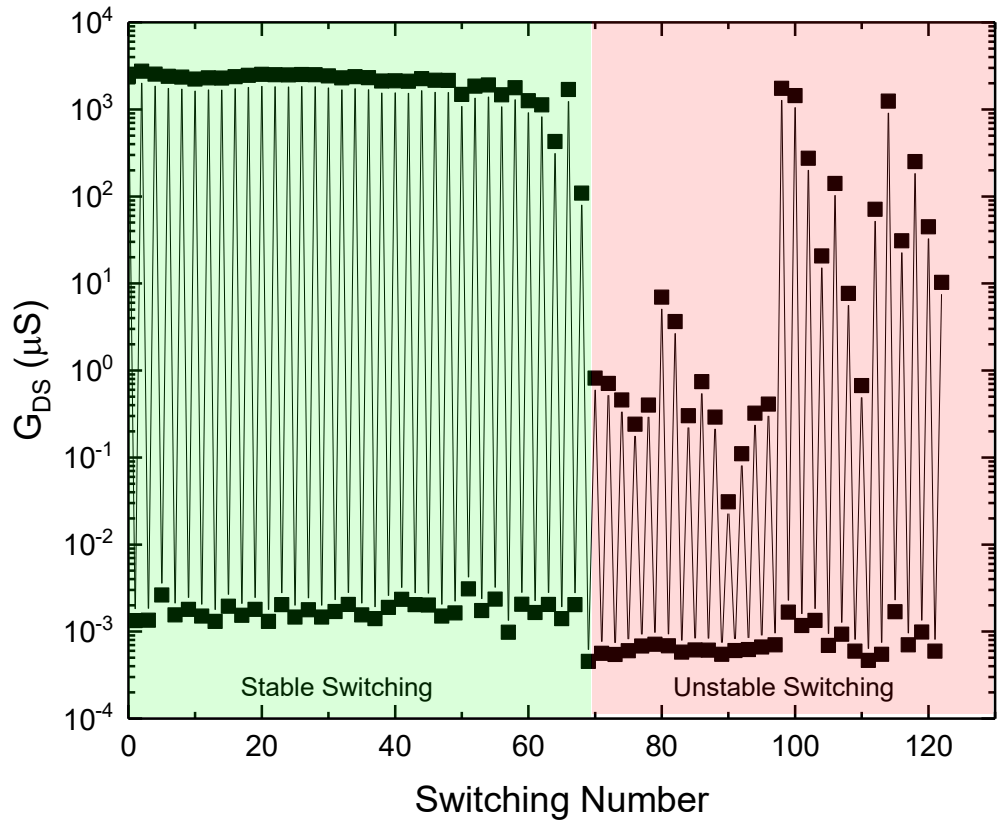

**Supplementary Figure 4.** Endurance test on a MoTe$_2$/PMN-PT (100) device (35 nm MoTe$_2$ thickness, W/L=1.4) with 50 nm Ni contact metal. Device switches between two different states upon application of +/-150 V in a stable configuration. Eventually after roughly 70 switches, the device evolves into non-stable switching. The reasoning behind this behavior is related to the change of the transfer characteristics as in Supplementary Figure 3. It is likely due to incomplete strain transfer through the thick MoTe$_2$ flake and can be further engineered for stability by reducing channel thickness.

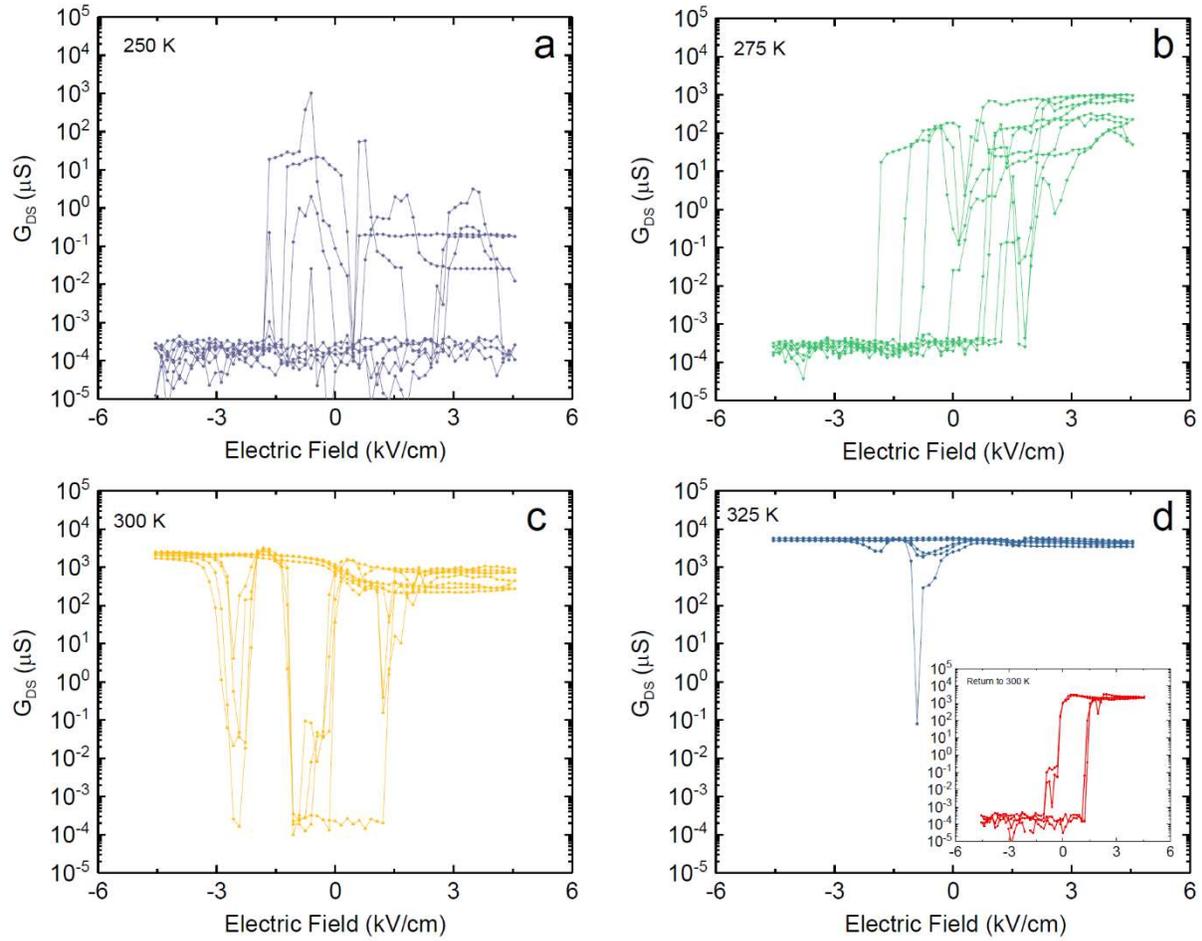

**Supplementary Figure 5.** Temperature dependent transfer characteristics on MoTe$_2$/PMN-PT (011) device (42 nm MoTe$_2$ thickness, W/L=3.92), separate to the data presented in the main paper and previous supplementary figure. Device was swept from 250 K to 325 K, and back to 300 K. The final unipolar device characteristics are presented in the inset of (d). Four full gate voltage sweeps are presented at each temperature, representing the steady state behavior of the device after two initial training sweeps (not shown for clarity). Device behavior at some temperatures continues to evolve with each progressive sweep showing a unique strain evolution within the device. Behavior presented here is also seen in data from the main paper, suggesting a temperature dependent phase transition between the semimetallic and semiconducting phases of MoTe$_2$, with the semiconducting phase favored at low temperatures.

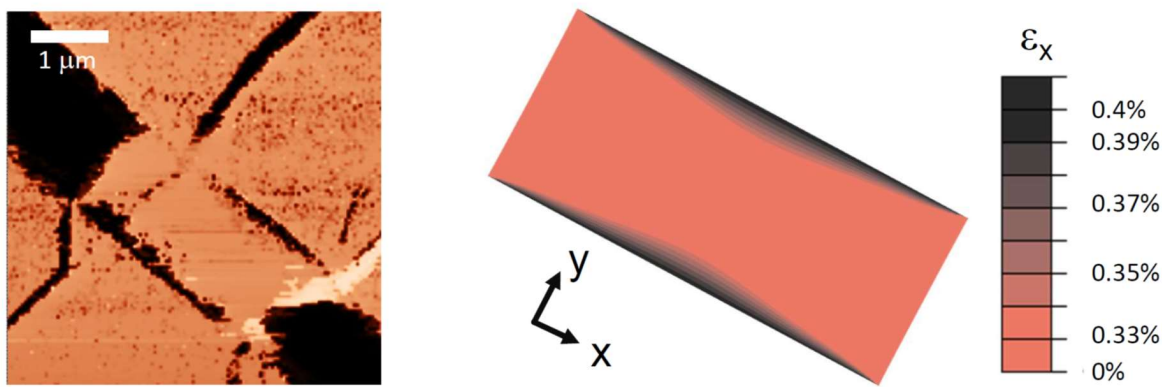

**Supplementary Figure 6.** Finite element analysis simulation of a rectangular channel device with contact pad induced strain clamped at 0.4%, presented with the CAFM image from Figure 3d reproduced. The characteristic features of the semiconducting region near the contacts are reproduced in the simulation. Simulations can provide an estimate as to how much strain is needed to seed the transition by matching the length scale of the observed semiconducting transition from CAFM images to the strain from simulations. From our data, it is estimated that the strain cutoff between the two phases is 0.33% tensile strain, which is represented in the simulation figure as a hard cutoff in color.

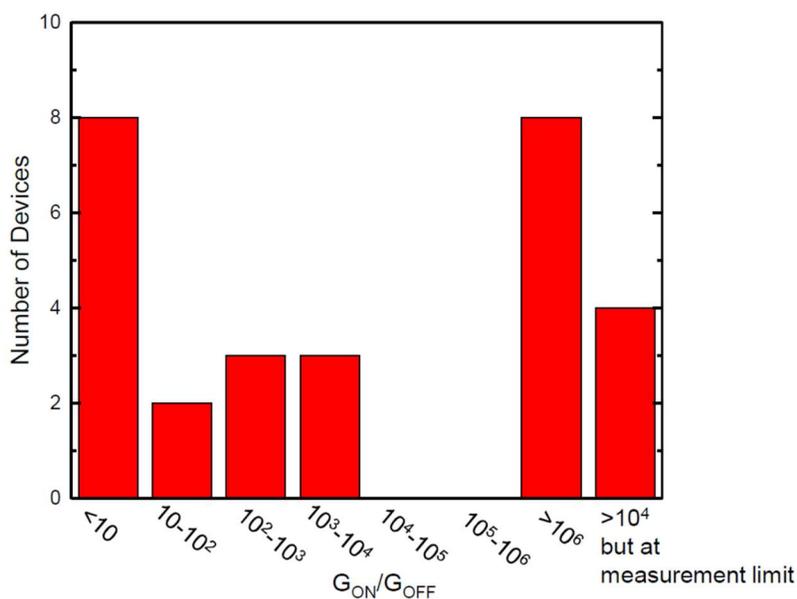

**Supplementary Figure 7.** Distribution of on-off ratios in all devices measured in the paper using Ni contact metals on PMN-PT (011) and (111) ferroelectric substrates. Devices typically entirely turn off or do not significantly change in conductivity since the 1T' phase acts as a short across the channel if the phase change to semiconducting does not result in complete channel cutoff. Variation is likely due to various uncontrolled factors like PMN-PT domain structure, or $MoTe_2$ crystal orientation. Four devices exhibited large on-off ratios (>$10^4$), but the measurement sensitivity at the time of characterization was not able to determine if the true on-off ratio was greater. Later characterization allowed for more sensitive measurement of the lowest conductivity state.

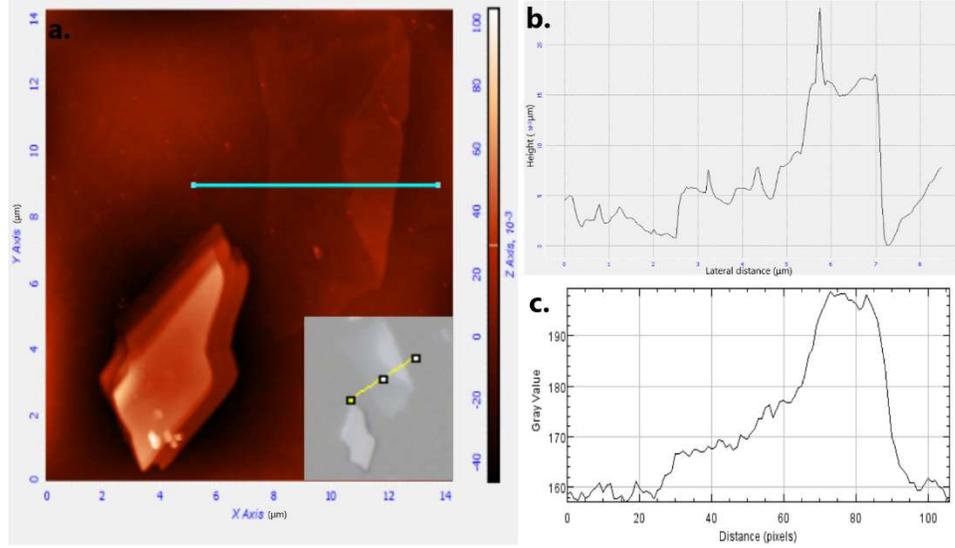

**Supplementary Figure 8.** Optical contrast measurements for MoTe$_2$ flake thickness determination. (a) Atomic force micrograph of two 1T'-MoTe$_2$ flakes exfoliated onto a PMN-PT substrate, inset is an optical micrograph of the same area (all optical micrographs taken of flakes were taken with same brightness and exposure time). (b) Height profile from the cross section in the AFM image from (a). (c) Contrast values of cross section in the optical micrograph presented in the inset of (a). Combinations of AFM and contrast measurements of various 1T'-MoTe$_2$ flakes on PMN-PT were used to determine contrast values for various thicknesses.

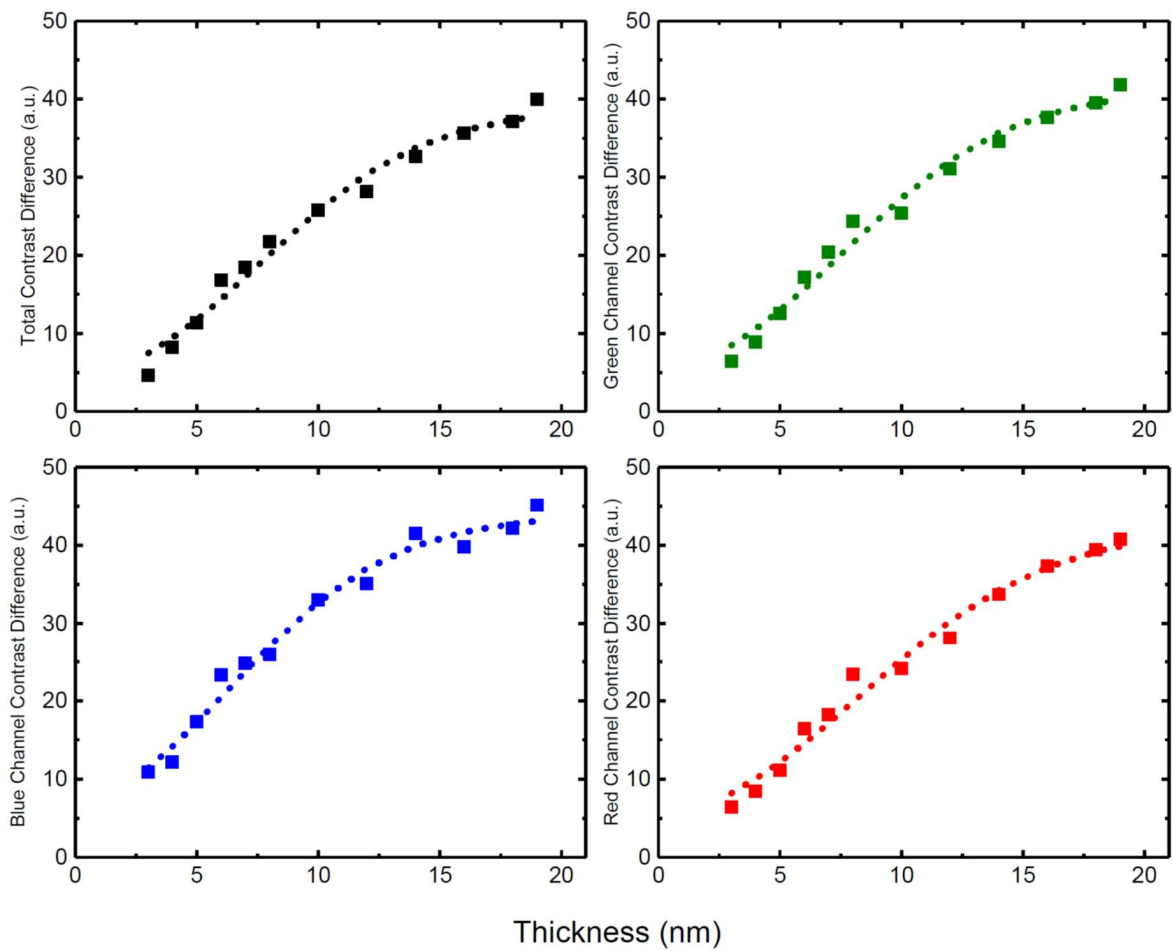

**Supplementary Figure 9.** Optical contrast calibration measurements using ImageJ image analysis software. The contrast of the substrate was found and then subtracted from the contrast value of the flake of interest. The contrast differences versus thicknesses were fitted to the Boltzmann function[3]. This function was then used to determine contrast values of various thicknesses (up to ~20nm). We did not use the function to extract thickness for any thicknesses larger than 20 nm since thicker flakes tended to reach the maximum contrast value shortly after 20 nm at our exposure level. All thicknesses determined from this method were then confirmed via AFM measurements. For flakes above 20 nm, AFM was used to extract channel thickness. Overall accuracy is determined to be roughly ±1 nm.

| Contact Material | Thickness | Radius of curvature on Kapton | Radius of curvature on Glass | Stress |
|---|---|---|---|---|
| Ni | 35 nm | 0.35 m | 7.08 m | 0.58 GPa |
| Ag | 50 nm | -1.02 m | -21.87 m | -0.20 GPa |

**Supplementary Table 1.** Calculated stress values of deposited contact metals through wafer curvature methods. Films were deposited on large (> 1 cm x 1 cm) pieces of Kapton and glass cover slides, with the radius of curvature of the substrate measured using contact profilometry post-deposition. Stress values were then calculated using the Stoney equation[4–6] using mechanical properties of Kapton.

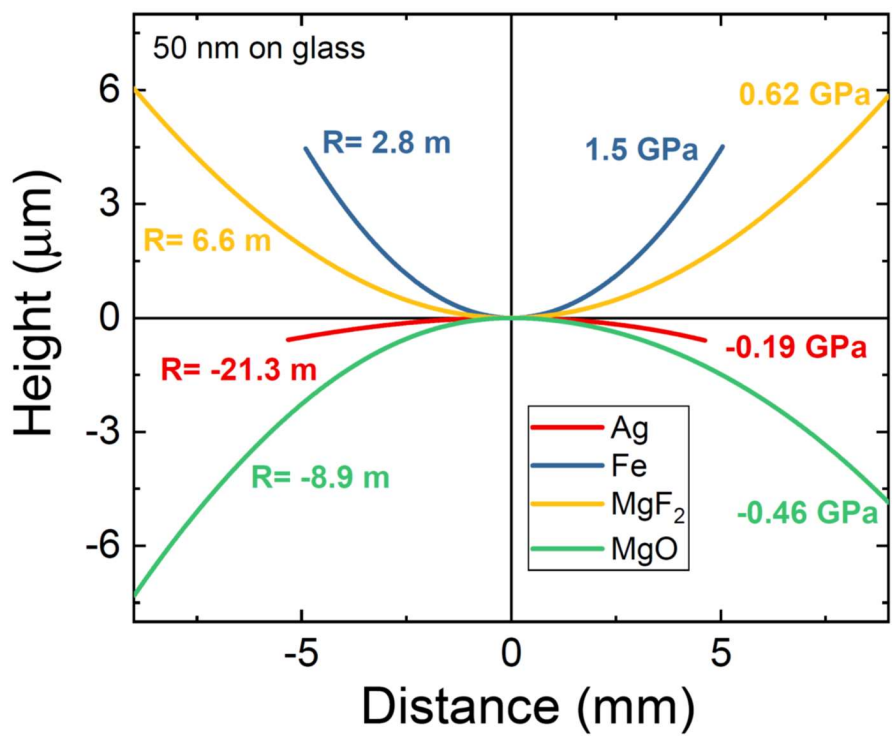

**Supplementary Figure 10.** Using the method described in Supplementary Table 1, thin film stress for e-beam evaporated 50 nm thin films of various materials were calculated using contact profilometry measurements of substrate curvature.

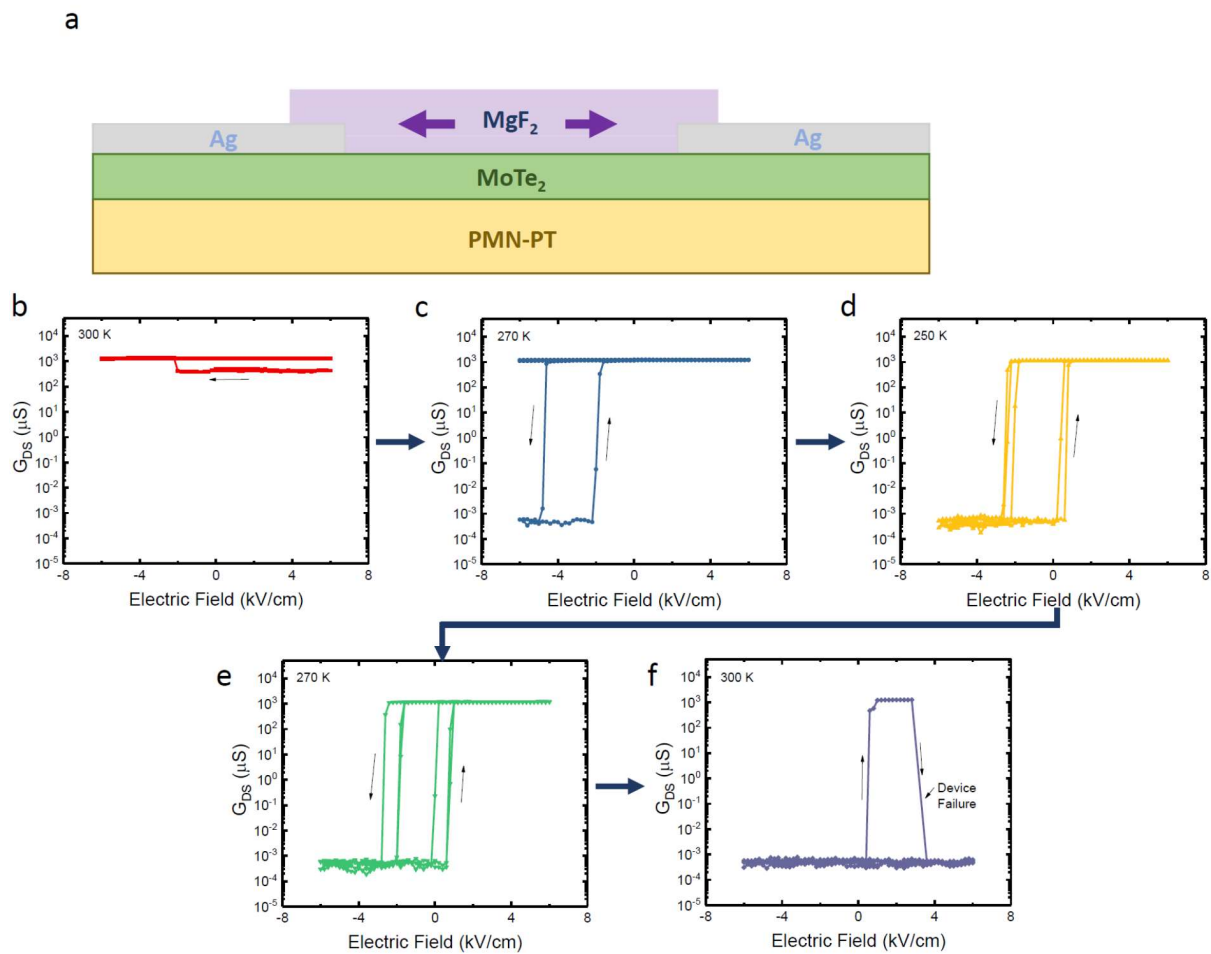

**Supplementary Figure 11.** Using an alternative straining mechanism, a device is patterned with an $MgF_2$ layer as a tensile stressor (0.62 GPa), with minimal stress from Ag contacts. Devices once again switch and behave similarly to devices with Ni contact metals. This suggests that chemical change transfer and other specific interactions between Ni and $MoTe_2$ are not the cause for our observed results and that strain is the dominant mechanism.

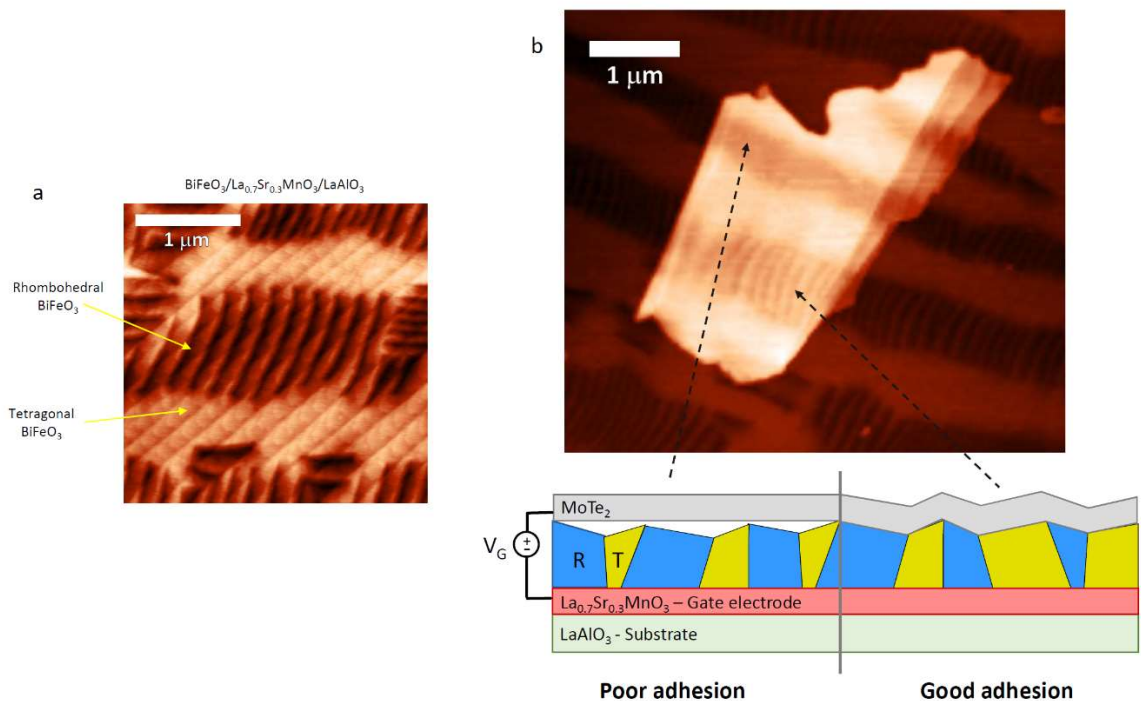

**Supplementary Figure 12.** Replacing ferroelectric PMN-PT with mixed-phase ferroelectric $BiFeO_3$ grown on a $LaAlO_3$ substrate with the coexistence of the rhombohedral (R) and tetragonal (T) structural phases[7,8], allows us to judge adhesion by observing local flake conformity to the substrate[9]. (a) AFM image of our epitaxially grown 150 nm $BiFeO_3$/5 nm $La_{0.7}Sr_{0.3}MnO_3$/$LaAlO_3$ heterostructure showing mixed phase $BiFeO_3$ exhibiting high quality growth with individual atomic terracing with both R and T phases present as noted from local corrugation. Samples were grown using ultra-high vacuum 90° off-axis RF sputtering. (b) AFM image of a $MoTe_2$ flake exfoliated on mixed phase $BiFeO_3$ showing regions which conform to the corrugation vs. regions that do not, a representative measure of the adhesion of the flake. Factors such as local surface roughness and flake adhesion may play a role in device reliability and endurance, which need to be systematically determined before the ultimate performance limit of the devices are known.

**Backside Raman for Phase-Identification**

To further examine the properties of the semiconducting region in our devices after phase changes have occurred, we perform backside Raman spectroscopy on MoTe$_2$ devices with Ni contacts fabricated on transparent double-side polished MgO substrates (0.25 mm thickness) in the same geometry as on PMN-PT substrates. By performing Raman measurements through the substrate, we are able to examine the phase-changed material underneath the Ni contact. We confirm the phase changes by first measuring Schottky contact behavior in the current-voltage characteristics. Supplementary Figure 13 shows that within the phase changed region two extra Raman peaks arise (127.3 & 143.2 cm$^{-1}$), that do not exist in either the pristine 1T' flakes or the unstrained parts of the channel. These peaks do not directly correspond to peaks associated with the pristine 2H semiconducting phase of MoTe$_2$ (A$_g$~173.3 cm$^{-1}$, E$_{2g}$~234.0 cm$^{-1}$). Peaks near 127 and 143 cm$^{-1}$ have been observed in various MoTe$_2$ samples from literature and have been associated with a widely varying degree of materials properties (semiconducting[10], semimetallic[11–13], optical transparency[14]…). The most common place where these additional peaks occur are in other phase-changed MoTe$_2$ samples where they appear regardless of the phase-change mechanism (strain[11] or laser induced[10,12,13]) and regardless of direction of phase change (2H→1T'[11–13] or 1T'→2H[10]). Recent studies on memristive MoTe$_2$ phase-change devices also show through detailed TEM studies that conducting metaphases between fully 1T' and fully 2H structure also exist in stable electronic devices when phase changes occur in MoTe$_2$[15]. Our hypothesis is that the additional peaks in our Raman measurements are associated with stable metaphases of MoTe$_2$ that exist between 1T' and 2H whenever a phase change mechanism occurs, which may be either semiconducting or semimetallic depending on the exact configuration of the metastable phase. This is supported by evidence from all previous MoTe$_2$ phase change studies, and from recent Raman studies that show laser induced phase-changed 2H-MoTe$_2$ relaxes into the thermodynamically stable version of 1T' upon annealing, with the 1T'-B$_g$ peak (163.4 cm$^{-1}$) appearing in addition to the 127/143 cm$^{-1}$ phase change peaks[13]. Further, we eliminate the possibility that the origin of these peaks are due to trivial effects such as defect formation. First, the unique Raman signature only occurs in well-bonded flakes with tensilely stressed contacts. This means damage from contact metal deposition or device processing cannot induce the same signature, and in devices with purposely low adhesion (exfoliated in humid environments) or debonded flakes (too much strain from the metal layer peels the flake away from the substrate) we only see the original 1T' Raman signature. We note that our Raman measurements show a **reduction** of these unique peaks upon repeated Raman measurement at 2 mW laser power (Supplementary Figure 14) that arises from low power laser heating, estimated to be in the 40-50° C range from literature[10]. This effect is seen in many samples and is the **opposite** behavior of all MoTe$_2$ laser phase patterning works[10,12], indicating heat will stabilize the 1T' phase (figure 2 from the main text), but not induce phase changes due to defect formation that require heating to 300-400° C at much higher laser powers. Further experimental and theoretical study on the exact nature of the metastable phases of MoTe$_2$ may be needed to find out the exact nature of both our phase-change device as well as all others in literature.

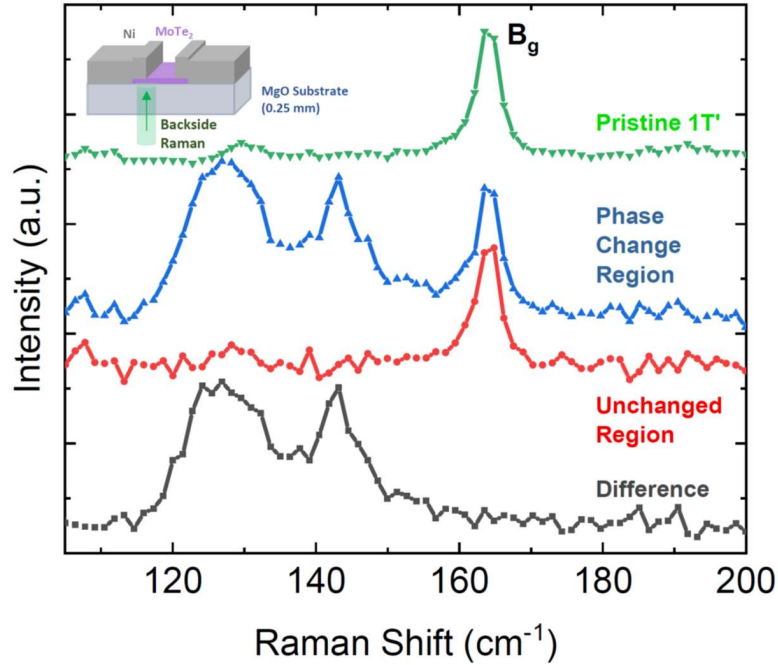

**Supplementary Figure 13.** Raman spectroscopy measurements taken through a transparent MgO substrate with the same patterned device as from previous PMN-PT experiments with Ni contacts. Devices were confirmed to start in the low-conductivity Schottky state before Raman measurement. Two regions of the same device are compared, one with significant additional Raman peaks (127.5 & 143 cm$^{-1}$) representing the signal from the phase changed MoTe$_2$. Concomitant 1T'-MoTe$_2$ signal is measured in the phase-changed region, due to either large laser spot size (1.5 μm) or a mixed-phase nature of the "changed" region. Pristine uncapped 1T'-MoTe$_2$ and the difference between the changed/unchanged region are also presented for comparison.

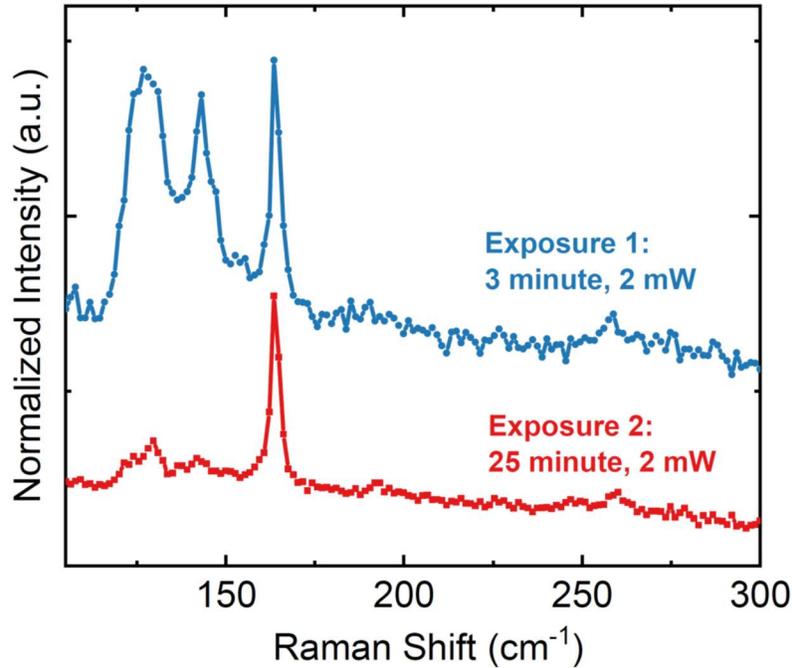

**Supplementary Figure 14.** Raman measurements of a phase-changed region using backside Raman microscopy under two exposures in the same region. The first exposure is reproduced from supplementary figure 13, and the second exposure is taken at the same location for a longer period. The longer exposure leads to more local heating (estimated to be 40-50° C from literature[10]) and reduces the size of the extra Raman features. This is the **opposite** behavior to if there were phase-changes due to defects or local laser heating[10,12]. Additionally, devices patterned for Raman with only **static** strain on MgO substrates, exhibit a transition from the high-conductivity to the low-conductivity state upon temperature cycling to 250 K and back to room temperature. This evidence aligns with strain driven phase changes, which are sensitive to temperature as seen from figure 2 of the main text but would not occur if any of the effects we observe in this work were related to $MoTe_2$ damage. Raman spectra between the two measurements were individually normalized to the background for comparison.

**Strain Memory Effect**

In the main paper, figure 1c shows strain measurement using strain gauges patterned directly on the ferroelectric, such that the underlying initial domain structure is undisturbed (the same situation as our $MoTe_2$ flakes when exfoliated). We use the same nickel thin film that is deposited for our contact metal and is separately calibrated for strain by depositing the gauge on flexible substrates. The measure of strain in the strain gauge on the ferroelectric is therefore effectively a measure of strain caused by our Ni contact metal in the $MoTe_2$ since the same material is used for both. We see that while initially the devices stay in the bipolar state, additional sweeping causes a transition to a unipolar strain state. This is further enhanced after temperature cycling from 300 K to 250 K to 330 K and back to 300 K. Since the strain measured in the nickel thin films is a direct measure of how much controllable strain we can apply to the $MoTe_2$ channel, we can see this as a direct experimental evidence of why our $MoTe_2$ transistors all eventually transition to unipolar. In this case we are not modulating just the strain on MoTe$_2$ using a ferroelectric, but we are also modulating the thin film stress from the nickel stress capping layer.

This type of transition to unipolar behavior is not unusual and is seen in many ferroelectric systems as a sign of internal bias field that occurs due to locking of frozen elastic dipoles[16,17]. This has been a long-studied phenomenon in ferroelectrics due to fatigue[18]. We hypothesize that the stress from the thin film causes elastic/electric dipoles in the ferroelectric system to become frozen upon sweeping, setting up an internal electric field bias. This effect has been seen in systems such as compressively strained thin film BaTiO$_3$ grown on various lattice mismatched substrates, where the coupled nature of the elastic and electric defect dipoles causes a large electric field bias in the ferroelectric polarization hysteresis curve[19]. Other ferroelectric systems where elastic/electric dipole orientation is related to external stress also exist[20–22].

When looking at other systems of thin films on PMN-PT (011) from literature, particularly among works where tensile or compressive magnetic thin films are deposited on a ferroelectric substrate, strain asymmetry is almost always present to some degree[23–27]. This also applies for systems where magnetic thin films on BaTiO$_3$ single crystal substrates as well[28–30], providing further evidence of stress induced dipole ordering that was presented in thin film BaTiO$_3$. These experimental findings from literature suggest that in addition to stress **in** BaTiO$_3$ itself causing defect dipole alignment[19], that thin film stress **on** a BaTiO$_3$ substrate can cause the same internal bias in BaTiO$_3$ as well. The actual evolution of thin film stress in our thin film stress capping layers is likely more complex as it may involve plastic deformation of the thin film layers as well. The exact microstructural evolution of thin film stress on ferroelectrics and the microscopic origin of ferroelectric internal bias in PMN-PT will need to be further studied to shed light on our experimental findings in our strain gauges.

To further highlight that a transition of the stress in the nickel layer is what causes our phase transition to change to unipolar, we include the actual temperature evolution of one nickel strain gauge on our PMN-PT (011) sample and compare it to the temperature evolution of the device presented in figure 2 of the main text (supplementary figure 15). We plot tensile strain in the negative-y direction to highlight the similarities of the strain gauge and the channel conductance data as we sweep electric field. In our 'strain bias' phase diagram from figure 4a in the main text, we know that tensile strain will cause the low conductivity state to exist. By choosing a point that we consider the "phase transition" point in the strain gauge data, we see a clear connection to the corresponding channel conductivity curve at the same point in temperature cycling. As we cycle the devices in temperature, both the strain in the strain gauge and point we choose as the transition point between semiconducting and semimetallic evolve with temperature, with the semiconducting phase being more stable at low temperatures and the semimetallic phase more stable at high temperatures. Our strain gauge data is just a representative evolution of strain in the device since it was measured on a separately patterned strain gauge and there is no effective way we can measure the strain on the device itself directly.

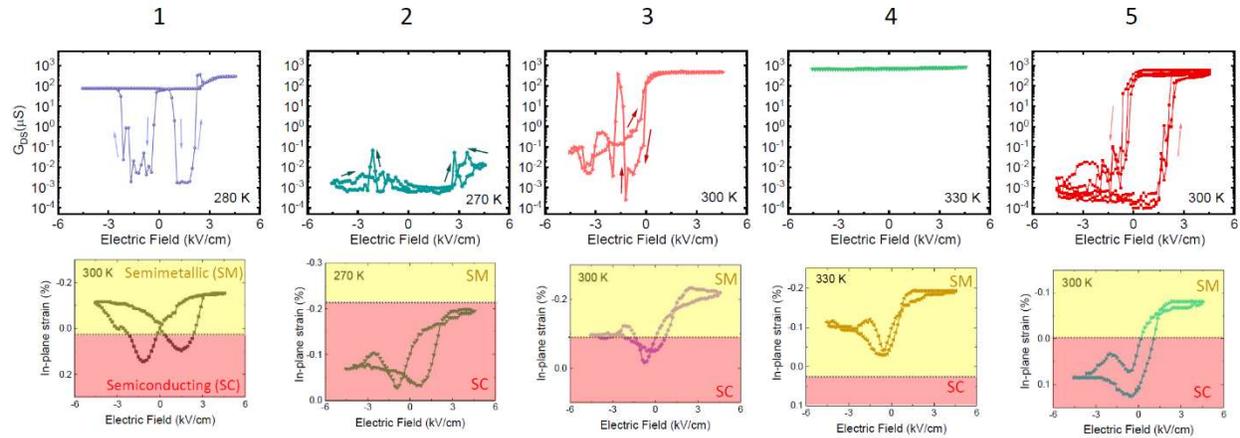

**Supplementary Figure 15.** Measurement of strain on a nickel strain gauge patterned on PMN-PT (011) with the same thin film used in the metallization step as the nickel contacts used in MoTe$_2$ strain transistor devices. Gauges are swept with applied electric field across the ferroelectric, and cycled in temperature from 300 K to 270 K to 300 K to 330 K to 300 K. The corresponding temperature evolution of the device presented in figure 2 of the main text is included, and a strain is chosen on each strain gauge curve to represent the "phase transition point" that results in the corresponding transfer curve. The bipolar curve at 280 K was chosen for "1" because it was the temperature where bipolar operation occurred, and is compared to the strain gauge data at 300 K where the strain is most bipolar.